# Integrated, ultrafast all-optical polariton transistors with sub-wavelength grating microcavities

Pietro Tassan[1,2,+], Darius Urbonas[1,+,#,*], Bartos Chmielak[3], Jens Bolten[3], Thorsten Wahlbrink[3], Max C. Lemme[3,4], Michael Forster[5], Ullrich Scherf[5], Rainer F. Mahrt[1], and Thilo Stöferle[1,#,*]

[1] IBM Research Europe – Zurich, Rüschlikon, Switzerland

[2] Photonics Laboratory, ETH Zürich, Zürich, Switzerland, Switzerland

[3] AMO GmbH, Aachen, Germany

[4] Chair of Electronic Devices, RWTH Aachen University, Aachen, Germany

[5] Macromolecular Chemistry Group and Wuppertal Center for Smart Materials & Systems (CM@S), Bergische Universität Wuppertal, Wuppertal, Germany

[+] These authors contributed equally to this work.

[#] These authors have jointly supervised this work.

[*] Corresponding authors: dar@zurich.ibm.com, tof@zurich.ibm.com

## Abstract

All-optical logic has the potential to overcome the operation speed barrier that has persisted in electronic circuits for two decades. However, the development of scalable architectures has been prevented so far by the lack of materials with sufficiently strong nonlinear interactions needed to realize compact and efficient ultrafast all-optical switches with optical gain. Microcavities with embedded organic material in the strong light-matter interaction regime have recently enabled all-optical transistors operating at room temperature with picosecond switching times. However, the vertical cavity geometry, which is predominantly used in polaritonics, is not suitable for complex circuits with on-chip coupled transistors. Here, by leveraging state-of-the-art silicon photonics technology, we have achieved exciton-polariton condensation at ambient conditions in fully integrated high-index contrast sub-wavelength grating microcavities filled with a π-conjugated polymer as optically active material. We demonstrate ultrafast all-optical transistor action by coupling two resonators and utilizing seeded polariton condensation. With a device area as small as 2×2 μm$^2$, we realize picosecond switching and amplification up to 60x, with extinction ratio up to 8:1. This compact ultrafast transistor device with in-plane integration is a key component for a scalable platform for all-optical logic circuits that could operate two orders of magnitude faster than electronic ones.

## Introduction

The clock speed of electronic circuits has been stagnant at a few gigahertz for almost two decades because of the breakdown of Dennard scaling[1], which states that by shrinking the size of transistors they can operate faster while maintaining the same power consumption. All-optical logic with the ambition to supplement electrical circuits, particularly for high-speed signal processing and tasks that are not well parallelizable[2] and therefore cannot fully benefit from the ongoing density scaling of electronic transistors[3], has been an active field of research for several decades[4–7]. On this journey, a large variety of device concepts has been explored, including optical bistability in semiconductor quantum wells and other materials[6], semiconductor optical amplifiers[8], resonators based on rings[9], discs[10] and photonic crystals[11], plasmonics[12], Mach-Zehnder interferometers[13], excitons[14] and single molecules[15]. While certain technologies enable operation at speeds that greatly exceed those of existing electronic circuits, achieving scalability for numerous interconnected all-optical gates continues to pose a significant challenge. This issue affects not only free-space architectures[16] but is also attributed to the typically low intrinsic nonlinearities of materials and the necessity for precise sizing and matching of optical resonators throughout a chip[17]. To boost the nonlinearities, some recent approaches utilize the strong light-matter interaction regime, where an optically active material is embedded in a microcavity such that the coupling rate between optoelectronic excitations (correlated electron-hole pairs, excitons) in the material and photons in the cavity exceeds the loss rates in the material and the cavity[18], effectively forming hybrid light-matter exciton-polariton quasi-particles. By using parametric scattering or resonant excitation, optical amplifiers[19] as well as transistors[20] and routers[21,22] have been realized at cryogenic temperature, and more recently with lead halide perovskites at room temperature[23].

With organics, in addition to the large exciton binding energy permitting exciton-polaritons at ambient conditions[24], the polariton formation can be strongly enhanced by matching the detuning of the polariton state from the exciton reservoir to the energy of molecular vibrational modes[25]. At sufficiently high excitation density, nonlinear polariton condensation occurs[26,27], resulting in coherent macroscopic occupation of a single polariton mode. Light inserted into the microcavity can seed and promote polariton condensation through bosonic stimulation, which can be harnessed for all-optical transistors and logic gates[28,29] operating at sub-picosecond speed and controlled with tens or less signal photons[30]. However, a major downside common to these exciton-polariton devices is their vertical cavity design with distributed Bragg reflectors (DBR), resulting in a polariton condensate wavevector that is essentially perpendicular to the substrate. Such vertical configuration necessitates the use of external free-space optics to route emitted light between the individual transistors, resulting in significant propagation delays on the order of hundreds of picoseconds that ultimately prevent the realization of scalable ultrafast circuits. Alternative geometries without DBRs have employed waveguides with facets[31] or microwires[32] to demonstrate amplification and logic functionality with polaritons in ZnO, but a scaling path towards more complex circuits has been missing.

Sub-wavelength gratings built from materials with high refractive index contrast provide an alternative to DBRs[33,34] and have already been used in a vertical microcavity structure to realize exciton-polariton lasing[35]. Fabricating a pair of high-contrast gratings (HCG) perpendicular to the chip substrate, Fabry-Pérot-like cavities can be created[36,37] where the cavity mode is parallel to the substrate, enabling scalable in-plane architectures. Remarkably, compared to polariton condensation with metal[38,39] and silicon[40] metasurfaces, where the condensate extends over hundreds of micrometers, the effective modal area in HCG cavities is only about 1 $\mu m^2$, enabling extraordinarily compact optical circuits with comparably high device densities, which is beneficial for lowering power consumption and high speed.

Here, we establish in-plane polariton condensation at room temperature by introducing compact integrated sub-wavelength grating cavities filled with an organic ladder-type polymer. With one HCG cavity generating the in-plane control signal that constitutes the input "seed" for a second "transistor" cavity, we demonstrate transistor action on a picosecond time scale. We observe up to 60x gain and a switching contrast exceeding 8:1 with an architecture that provides a scalability path through the planar on-chip nature of the signal routing and the compact device.

## Results

### On-chip polariton condensation

As illustrated in Fig. 1a, the device comprises two HCG mirrors that provide lateral light confinement, and a polymer serving as both active material and guiding layer with vertical light confinement via total internal reflection, enabling in-plane propagation. The diameter and pitch of the nanostructured silicon pillars are designed to yield high in-plane reflectivity for transverse-electric light polarization at perpendicular incidence to the grating (see Supplementary Fig. 1). The wide reflectivity maximum in spectrum and configuration space ensures broadband reflectivity and excellent robustness against fabrication errors. The HCG microcavity is fabricated on a silicon-on-insulator wafer using state-of-the-art silicon photonics technology and subsequent deposition of the ladder-type π-conjugated polymer (MeLPPP) and the encapsulation layer (see Methods and Supplementary Fig. 2). Photonic simulations show the optical mode confinement within the device (Fig. 1c-e), resulting in a cavity quality factor of around $Q \sim 400$ for a wavelength of 490 nm, despite the strong optical absorption of the silicon constituting the HCGs. Notably, this $Q$-factor is sufficiently high to enable strong light-matter interaction and exciton-polariton condensation[26], while still being low enough to ensure sub-picosecond polariton lifetimes and to relax the fabrication tolerance requirements for matching resonator wavelengths across the chip.

We excite the structure from the top using a pulsed laser (see Methods). At low pump fluence, the intensity of the emission increases linearly (Fig. 2a, top panel), and a Lorentzian-shaped spectral peak with ~7 meV full-width at half-maximum (FWHM), consistent with the calculated cavity $Q$-factor, is observed along with a broad photoluminescence background. With increasing pump fluence, above threshold ($P_{th}$ = 39 μJ cm$^{-2}$) the emission peak acquires a Gaussian shape (Fig. 2a, inset and Supplementary Fig. 3a) and undergoes a linewidth narrowing down to ~1 meV. Furthermore, a continuous blue-shift up to 5 meV from its original photon energy at a pump fluence of 90 μJ cm$^{-2}$ is observed (Fig. 2a, bottom panel). To confirm that these observations are indeed signatures of polariton condensation, as already reported in vertical DBR cavities with the same active material and similar $Q$-factor[26], we assess the presence of strong light-matter coupling in these structures by measuring a manifold of devices where the cavity length $L$ is systematically varied, similar to the wedge-tuning in vertical DBR cavities[41,42]. Plotting the resonance energies versus the cavity length $L$, we observe the characteristic bending of the lower polariton branch in the strong light–matter interaction regime, that is supported by comparing to simulations of the weakly-coupled and the strongly-coupled polymer-filled cavities (Fig. 2b,c) where the resonance energies change with a distinctly different slope with $L$. Fitting with a coupled-oscillators model (Supplementary Fig. 4) yields a Rabi splitting of $2\Omega = 327 \pm 48$ meV for an excitonic fraction of 50% at $L$ = 1.7 μm. The large Rabi splitting and high excitonic fraction, compared to the values found in vertical DBR cavities where polariton condensation with MeLPPP was observed[26] ($2\Omega \sim 120$ meV and 20% exciton fraction), results from the significantly increased amount of polymer material in the cavity that is, however, partly offset by the reduced cavity finesse. Notably, for microcavities with MeLPPP, polariton condensation occurs only when the lower polariton branch is within ±20 meV the energy of the exciton reservoir (2.714 eV) minus the energy of the strong vibronic transition of the polymer[30] (200 meV), where vibron-enhanced relaxation is prevalent[43]. The above-threshold excitation fluence emission of MeLPPP is reasonably stable, exhibiting only few percent photodegradation after several hours-long excitation (Supplementary Fig. 3b). This could be potentially much further extended together with a further reduced threshold by changing the pump photon energy from 3.1 eV closer to the exciton resonance, resulting in less excess energy and higher absorption in the MeLPPP[30].

Introducing a curvature to the HCGs allows us to engineer the modal structure of the microcavity. Gratings with a Gaussian shape, besides increasing the light confinement along the grating, lead to the discretization of the transversal modes (Supplementary Fig. 5). The different transversal orders can be clearly identified from the real-space emission patterns of the polariton condensates, which show the parasitic vertically scattered light of the in-plane mode by the HCGs. Thus, by engineering the cavity length and curvature, we can ensure single mode polariton condensates with well-defined polarization (Supplementary Fig. 6), an important feature for controlled coupling of multiple HCG resonators.

For scaling to circuits with many devices, it is crucial that the resonance frequency of the fabricated HCG cavities is well controlled, and deviations stay within the linewidth of the resonators. The fact that the resonance energies of more than three hundred HCG cavities line up almost perfectly in the study of the mode structure (Supplementary Fig. 5), suggests good homogeneity and precision of the fabrication processes. Moreover, we studied the statistical distribution of polariton condensate emission energies of more than 30 nominally identical fabricated cavities (Supplementary Fig. 7), resulting in a standard deviation of < 0.5 meV, much below the cavity linewidth (FWHM ~ 7 meV). Hence, even accounting for blue-shift of the condensate emission, this fulfils a pivotal condition for device cascadability where the output of one cavity serves as input of another one.

To realize ultrafast devices, it is important to assess the dynamics of the polariton condensation. By exciting a single HCG cavity above threshold (~1.2 $P_{th}$) with a femtoseconds-pulsed laser and interfering the real-space images by means of a Michealson interferometer with a retroreflector mounted in one arm, we measured temporal and spatial coherence (Supplementary Fig. 8). The coherence time of the condensate, which is directly related to its lifetime, shows a FWHM ~270 fs, providing an estimate for the ultrafast pulse duration generated as HCG cavity output. Furthermore, we performed an experiment with a bi-injection scheme where we excited a single HCG cavity with two subsequent pulses of the same fluence and same duration (~150 fs), while precisely tuning the time delay between them. When the fluence of a single pulse is below threshold, the polariton condensation process with its concomitant increase in emission is triggered only when the second pulse arrives within about 1 ps (Supplementary Fig. 9), which we therefore attribute to an intrinsic relaxation time, consistent with modelling and experimental results with MeLPPP polaritons in vertical cavities[29,44]. For larger time delays, a much longer lifetime component around 22 ps is observed that corresponds to the lifetime of the excitons. Hence, from the time scale of the initial fast drop, the maximum repetition rate of the devices can be inferred, and the height of the long tail versus the initial peak gives a limit for the extinction.

**In-plane transistor action**

Next, we study a configuration comprising two cavities with curved HCG mirrors, having the same resonance energy and being spatially separated by a 3 μm gap. We excite independently each cavity with an ultrafast pulse of controlled fluence and a variable time delay $\Delta t$ between the two pulses (Fig. 3a). In this scheme, we use the output photons of one cavity, representing the control gate or "seed", as input for the other cavity, representing the "transistor". The polymer layer, that covers both structures as well as the space between them, serves as a waveguide with vertical confinement realized through total internal reflection due to the polymer's higher refractive index than the $SiO_2$ box layer and air cladding below and above, respectively. This vertical confinement, combined with low lateral divergence, allows the emission from the seed cavity to couple into the same mode of the transistor cavity after only few femtoseconds propagation due to the short distance between them. We estimate a coupling efficiency of ~84%, obtained from the ratio of the integrated intensity of parasitically scattered light from the transistor output mirror and the transistor input mirror when the seed cavity is pumped but the transistor cavity is not excited.

Real space images of the emission (Fig. 3b) indicate that the transistor can still achieve polariton condensation when exciting it below threshold ($P_{transistor}$ = 0.8 $P_{th}$), but only if the seed cavity is excited above its condensation threshold ($P_{seed}$ = 1.2 $P_{th}$) slightly before the transistor cavity with a negative seed–transistor pump pulse time delay around $\Delta t$ = -1 ps. We record the output intensity of the transistor cavity by collecting the emission spectrum only from the transistor HCG mirror that is on the far side from the seed, i.e. the transistor output. When we plot the emission intensity versus the excitation fluence (Fig. 3c), we find that the condensation threshold is reduced only when the gating signal from the seed is present and arrives with the aforementioned timing, but not when the seed cavity is excited later ($\Delta t = +1$ ps).

To map the temporal dynamics, we vary both the transistor excitation fluence and the time delay $\Delta t$ between the excitation pulses for the seed and the transistor cavity, for a fixed seed excitation fluence of $P_{\mathrm{seed}} = 1.5\ P_{\mathrm{th}}$ (Fig. 3d). Comparing the emission intensities, we observe that signal amplification (Fig. 3e) in the transistor only occurs when the seed cavity is excited ~1 ps before the transistor cavity. This optimal timing condition originates from the time required for the polaritons in the seed cavity to condense, as mentioned above in the two-pulses experiment with a single HCG cavity. Furthermore, we observe a significant threshold reduction of the transistor cavity (Fig. 3e), which for this seed fluence is about 15%, but can reach more than 30% with stronger seed ($P_{\mathrm{seed}} = 2\ P_{\mathrm{th}}$).

As both seed and transistor cavities are nominally identical, they can switch roles. When we detect the light on the HCG mirror of the seed cavity on the far side of the transistor cavity, we observe the mirrored temporal dynamics compared to Fig. 3d (Supplementary Fig. 10), with a higher background due to the above-threshold excitation of the seed cavity ($P_{\mathrm{seed}} = 1.2\ P_{\mathrm{th}}$). This demonstrates that the excitation pulse sequence determines the directional flow of the signal on the chip, suggesting that scalable circuits comprising multiple successive transistors with signal routing free from complications arising from back-action can be developed by utilizing customized excitation sequences. Furthermore, in contrast to vertical geometries[28,29], which need to reroute the off-chip transistor output back onto the chip, thereby incurring hundreds of picoseconds latency, the propagation time of the photons between different integrated cavities here is on the order of tens of femtoseconds.

**Transistor metrics**

We extract key transistor metrics from the emission spectra and their temporal dynamics. When comparing the emission spectrum at the transistor output without and with applying the excitation to the transistor cavity (Fig. 4a), we observe that the amplified peak faithfully reproduces the input from the seed, only negligibly changing its energy and spectral shape. Likewise, comparing spectra from the transistor output without and with seed excitation shows that the output spectrum remains essentially unchanged apart from the intensity change (Fig. 4a), allowing the extraction of the extinction ratio (on/off) of the transistor. The absence of significant spectral distortion is a fundamental requirement for the cascadability of the devices. We plot the transistor metrics as a function of transistor excitation fluence (Fig. 4b), showing that for a given seed excitation fluence ($P_{\mathrm{seed}} = 1.5\ P_{\mathrm{th}}$) the amplification increases with the transistor excitation up to a value of 43, and the extinction reaches a ratio of 8:1 near the threshold. Varying the seed excitation fluence, and thereby the intensity of the gating input to the transistor, we find that the signal amplification increases for smaller input signals (Fig. 4c), reaching 60 for the smallest inputs that we could produce without losing polariton condensation in the seed cavity. At this condition, where the seed pumping fluence is just above the seed cavity condensation threshold, also the switching energy is minimal, which is around ~2.9 fJ.

Furthermore, to assess the temporal robustness of the transistor amplification, we investigate how the amplification dynamics change with transistor excitation fluence (Fig. 4d). We find that the time delay between the seed and the transistor excitation where the maximum amplification is achieved slightly shifts from -0.9 ps to -1.4 ps with increasing transistor excitation. This can be understood considering that for stronger transistor excitation, the spontaneous polariton condensation is generally more dominant, and therefore the seed light must be injected early enough into the cavity to stimulate scattering into the seeded mode and suppress spontaneous condensation.

## Discussion

We realized optically-pumped room-temperature polariton condensates that, in contrast to previous non-scalable architectures, feature a single micrometer-sized condensate mode that is parallel to the chip surface, enabling integrated planar circuits. By cascading two resonators in-plane, we demonstrated all-optical ultrafast transistor action with femtojoule-level input signals and signal amplification of up to a factor of 60 that can enable large fan-out and provide scalability beyond this basic building block. Harnessing silicon photonics nanofabrication technology and exploiting the outstanding engineerability of the HCG cavities[45], we expect our approach to be scalable to combine several transistors to logic gates[28] and more complex logic circuits. Using waveguides with either HCGs[46] or silicon nitride co-integration[47,48] would allow routing the signals between distant transistors as well as the excitation pulses for the polariton condensates, opening the door towards a platform for all-optical logic circuits that may operate significantly faster than electronic ones.

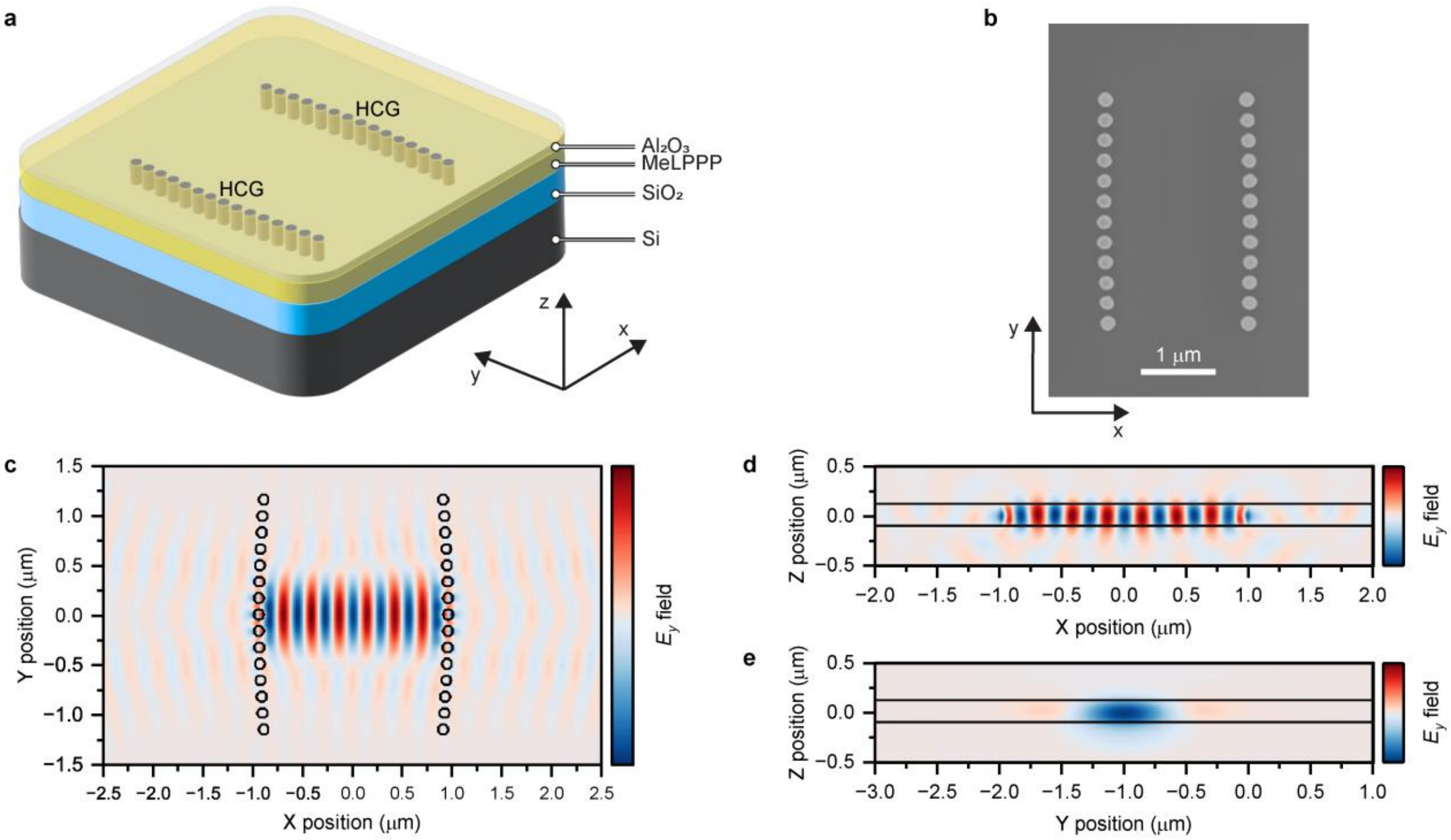

**Fig. 1 | Integrated high contrast grating cavity. a**, Sketch of integrated high contrast grating (HCG) cavity embedded within the active polymer (MeLPPP) and encapsulation layer ($Al_2O_3$). **b**, Scanning electron microscopy image (top view) of a fabricated device without polymer layer. **c-e**, Different cross-sectional views of a three-dimensional finite-difference time-domain (3D FDTD) simulation revealing the field distribution of the resonant mode in the cavity. Overlaid in black are the outlines of the structure.

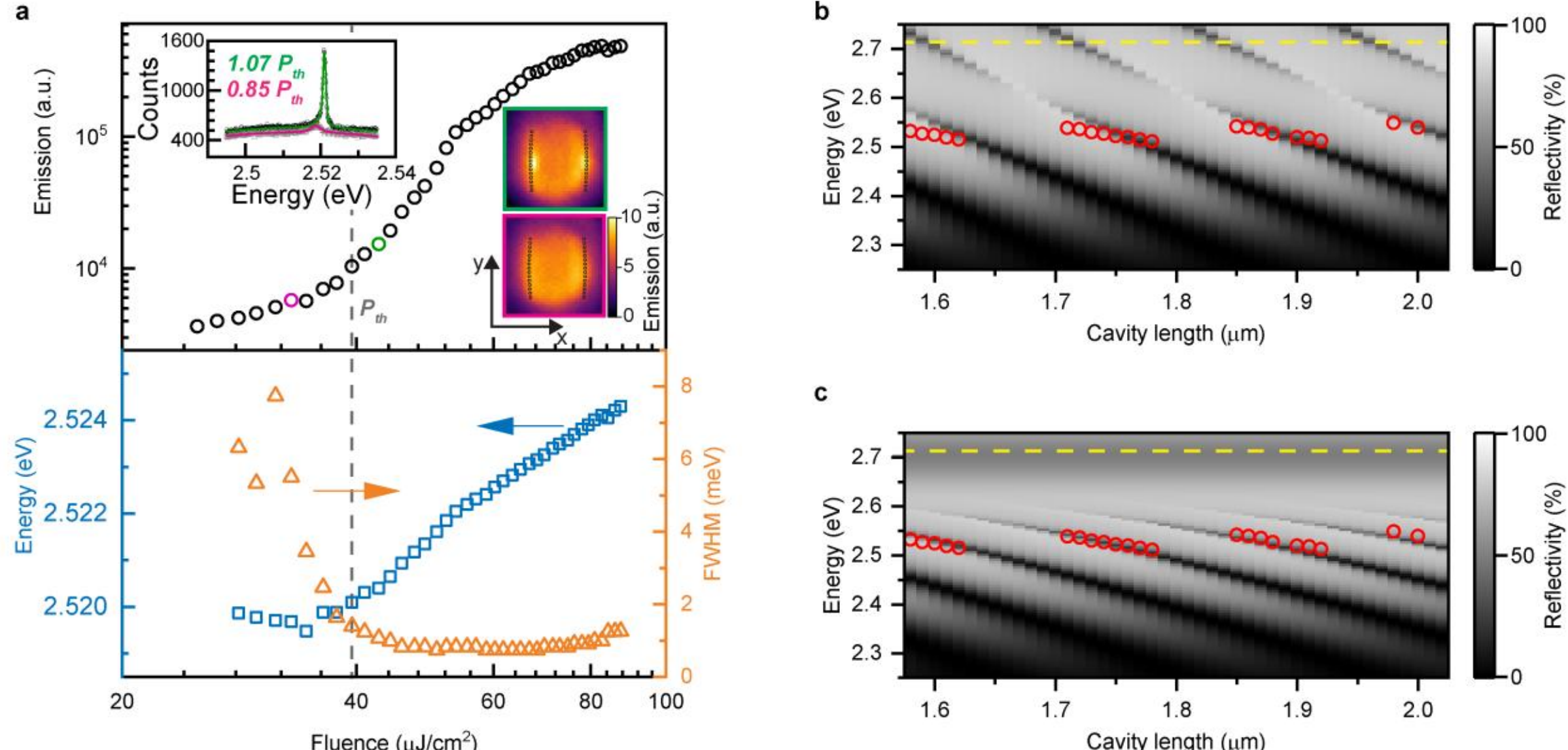

**Fig. 2 | Strong light–matter interaction regime and polariton condensation. a**, The top panel shows the emission as a function of excitation fluence. This light-in light-out characteristic exhibits a nonlinear increase above the threshold ($P_{\mathrm{th}} \sim 39$ μJ cm$^{-2}$, dashed line) and saturation at higher fluences. The top left inset shows the emission spectra below threshold (gray data points) and above threshold (black data points), and their fitted spectra (magenta and green, respectively). The right insets are showing the real-space emission images (4.2 μm × 4.2 μm) of the cavity when pumped below and above threshold, corresponding to the fluences with the magenta and green data points, respectively. The bottom panel displays the center energy of the emission peak (blue squares) and the width (orange triangles) as a function of excitation fluence, showing a sudden narrowing at the threshold and a continuous blue-shift above. **b**, Reflectivity (grey scale) obtained from a two-dimensional (2D) rigorous coupled wave analysis (RCWA) calculation as a function of cavity length in the weak coupling regime with refractive index $n = 1.86 =$ const. and $k = 0$; the different longitudinal modes appear as dark stripes in the grey scale data. Experimentally measured energies of the polariton condensates are shown as red circles, clearly not matching the slope of the weak-coupling theory. The exciton energy is represented with a yellow dashed line at 2.714 eV. **c**, Simulated reflectivity (RCWA) as a function of cavity length for cavities in the strong-coupling regime (grey scale), where the full refractive index dispersion $n = \mathrm{n}(\lambda)$ and $k = \mathrm{k}(\lambda)$ is used, as obtained from ellipsometry of the polymer layer. The experimentally measured energies of the polariton condensates are overlaid as red circles. To account for the 2D nature of the simulation, which neglects the vertical guiding layer structure that results in a lower effective refractive index, the theoretical curves are shifted slightly (+0.07 μm in cavity length) to achieve a good match with the experiment.

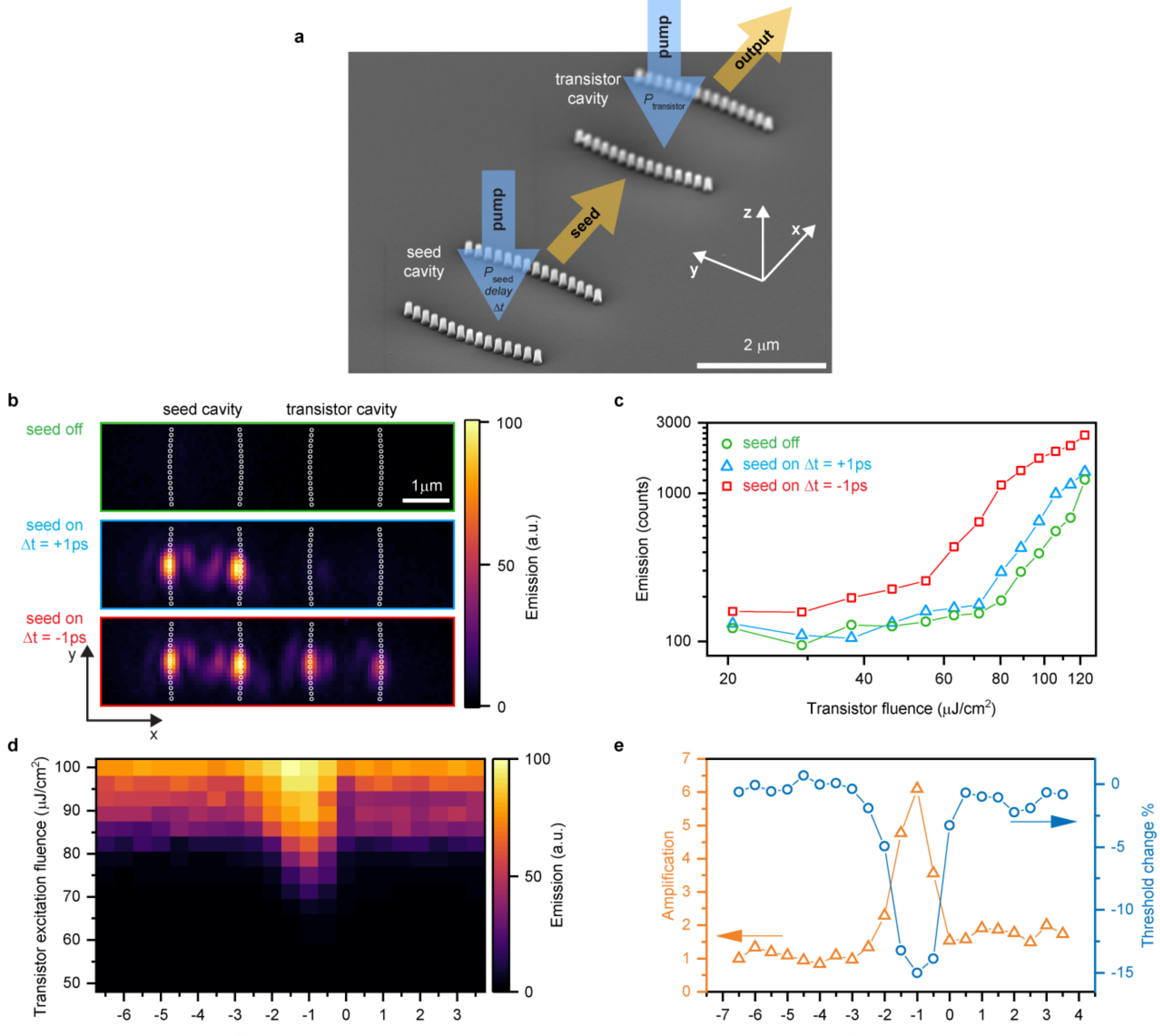

**Fig. 3 | Ultrafast transistor action. a**, Illustration where a scanning electron microscopy image of the seed and the transistor cavity without polymer layer is overlayed with arrows for the excitation beams and the output of the seed cavity flowing to the transistor cavity. **b**, Real space images of the emission with excitation of the transistor cavity at $P_{transistor} = 0.8\ P_{th}$. In the top green panel, the seed cavity is not excited. In the middle blue panel, it is excited with $P_{seed} = 1.2\ P_{th}$ at a time delay of $\Delta t = +1$ ps after the excitation of the transistor cavity. In the bottom red panel, $P_{seed} = 1.2\ P_{th}$ at $\Delta t = -1$ ps before the excitation of the transistor cavity. **c**, Integrated intensity of the emitted spectrum collected from the output mirror of the transistor cavity shown as a function of transistor excitation fluence when the transistor cavity is excited at $P_{transistor} = 0.8\ P_{th}$ and the seed cavity is not pumped (green), is excited with $P_{seed} = 1.2\ P_{th}$ at a time delay of $\Delta t = +1$ ps (blue) and with $P_{seed} = 1.2\ P_{th}$ at a time delay of $\Delta t = -1$ ps (red). **d**, Map of the transistor output emission intensity versus transistor excitation fluence and time delay between seed and transistor excitation for $P_{seed} = 1.5\ P_{th}$. **e**, Threshold reduction of the transistor cavity (blue) and Amplification of the seed (orange) as a function of seed–transistor excitation time delay.

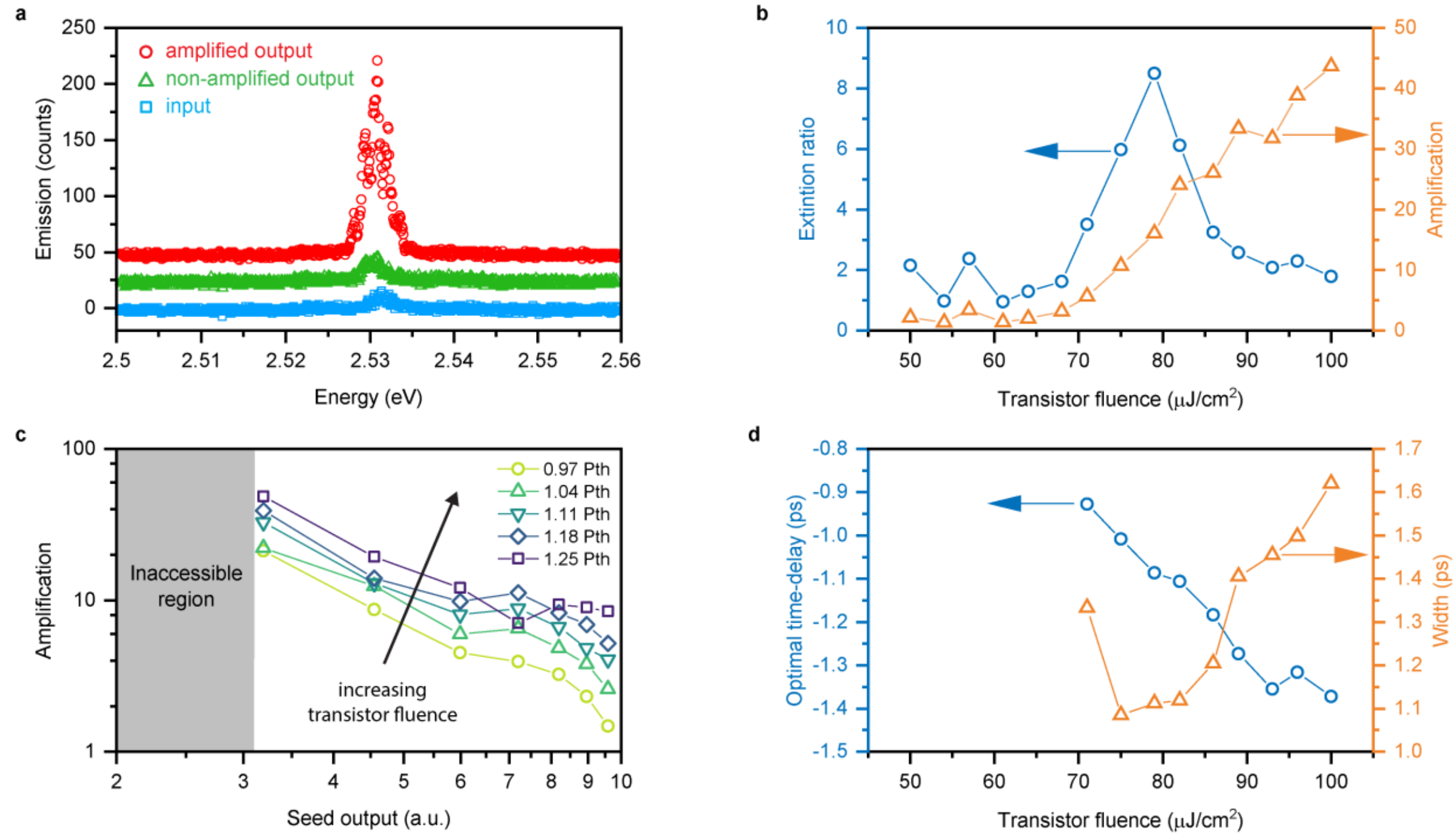

**Fig. 4 | Spectra and transistor performance. a,** Comparison between the input signal (blue), the transistor output spectrum without amplification (green) and the amplified signal (red), offset vertically by 25 counts each. The spectra have been collected at the transistor output mirror when the seed is excited with $P_{\mathrm{seed}}$ = 1.2 $P_{\mathrm{th}}$ and the transistor with $P_{\mathrm{transistor}}$ = 0.6 $P_{\mathrm{th}}$ (blue) and the transistor with $P_{\mathrm{transistor}}$ = 1.4 $P_{\mathrm{th}}$ at $\Delta t$ = +1 ps (seed pumped after transistor, green) and $\Delta t$ = -1 ps (seed pumped before transistor, red), respectively. **b**, Extinction ratio (blue) and amplification (orange) of the transistor versus transistor excitation fluence. **c**, The transistor amplification varies as a function of the intensity of the input signal, i.e. the seed cavity output which is related to the seed excitation fluence, and with the transistor excitation fluence. In the inaccessible region, the seed cavity excitation is too weak to induce polariton condensation in the seed cavity thereby preventing generation of a short input pulse for the transistor. **d**, Optimal time delay (blue) and width of the amplification time-window (red) as function of the transistor excitation fluence.

# Methods

## Device fabrication

We have fabricated the HCGs on silicon-on-insulator (SOI) wafers with 220 nm top-Si thickness and 3 μm buried oxide using high-resolution silicon photonics technology, where electron beam lithography (Raith EBPG 5200) exposes hydrogen silsesquioxane (HSQ) resist with 200 nm layer thickness. After development with tetramethylammonium hydroxide (TMAH), the structures have been transferred by Cl-based reactive ion etching (RIE) into the top-Si layer. This results in a grating post diameter of 110 nm and a pitch of 170 nm. Methyl-substituted ladder-type poly(p-phenylene) (MeLPPP; *Mn* = 31,500 (number averaged molecular weight), *Mw* = 79,000 (weight averaged molecular weight)) was synthesized as described elsewhere[49]. MeLPPP is dissolved in toluene and spin-coated on the fabricated HCG structures, resulting in approximately 220 nm layer thickness, verified by profilometry and ellipsometry. A 20 nm-thick encapsulation layer of $Al_2O_3$ is evaporated on top for protection against photodegradation.

## Photonic simulations

For the computation of the HCG reflectivity and the spectra of the empty cavity and filled with polymer, we perform rigorous coupled wave analysis (RCWA) using a freely available software package[50]. For the three-dimensional finite-difference time-domain (3D FDTD) simulations, we use the commercial software ANSYS Lumerical. For both calculations, we use the complex refractive index of the polymer material obtained from variable-angle spectroscopic ellipsometry measurements (Woollam VASE) of a test polymer layer on a silicon wafer.

## Optical characterization

We mount the 20 x 20 mm chips, each containing hundreds of HCG cavities, on an *XYZ* nanopositioning stage in ambient conditions. Ultrafast excitation light pulses of 400 nm wavelength, 150 fs pulse duration and 1 kHz repetition rate are generated from a frequency-doubled regenerative amplifier that is seeded by a mode-locked Ti:sapphire laser. For the cavity detuning experiments with single HCG cavities, we insert the pulsed light into a photonic crystal fiber that stretches the pulse duration to >5 ps and provides a near-Gaussian beam at its output. The light is focused through a microscope objective (Mitutoyo Plan Apo 100X, NA=0.7) to an approximately Gaussian spot size of ~3 μm $1/e^2$ diameter on the sample. For the experiments on the single cavity condensate dynamics and with multiple HCG cavities, we split the beam from the laser system and control the power and relative delay between the two beam paths before jointly inserting them into the same objective. Excitation fluences are controlled by spatially moving metallic gradient filters on linear stages. For detection, we collect the emitted light through the same objective and separate it with a dichroic beam splitter and a long pass filter from the excitation light. We route the emitted light with a 50:50 beam splitter to a camera and a monochromator with liquid nitrogen-cooled detector. The spectrometer collects the light via a multimode fiber with 10 μm diameter from an effective detection spot of ~1 μm diameter on the sample for the transistor measurements and 100 μm fiber diameter with detection spot of ~10 μm diameter for the other measurements.

## Data availability

Data supporting the findings of this study are available from the corresponding authors upon reasonable request

## Acknowledgements

We thank the team of the IBM Binnig and Rohrer Nanotechnology Center, Daniele Caimi and the Quantum Photonics team for support with the sample fabrication and Pavlos Lagoudakis for stimulating discussions. We acknowledge funding from EU H2020 EIC Pathfinder Open project "PoLLoC" (Grant Agreement No. 899141) and EU H2020 MSCA-ITN project "AppQInfo" (Grant Agreement No. 956071).

## Author contributions

P.T. performed the experiments and analysed the experimental data, supported by D.U.. B.C., J.B., T.W. and M.C.L. fabricated the silicon photonic structures. M.F. and U.S. synthesized the polymer. P.T. deposited the polymer and encapsulation layer, and D.U. characterized the layers. D.U. performed the photonic simulations supported by T.S. and designed the sample, supported by P.T.. R.F.M., T.S. and D.U. conceived the concept and supervised the work. P.T., R.F.M., D.U. and T.S. wrote the paper with contributions from all authors.

.

## Competing interests

The authors declare no competing interests.

## Additional information

**Correspondence and requests for materials** should be addressed to Darius Urbonas (dar@zurich.ibm.com) and Thilo Stöferle (tof@zurich.ibm.com).

# SUPPLEMENTARY INFORMATION

# Integrated, ultrafast all-optical polariton transistors with sub-wavelength grating microcavities

Pietro Tassan[1,2,+], Darius Urbonas[1,+,#,*], Bartos Chmielak[3], Jens Bolten[3], Thorsten Wahlbrink[3], Max C. Lemme[3,4], Michael Forster[5], Ullrich Scherf[5], Rainer F. Mahrt[1], and Thilo Stöferle[1,#,*]

[1] IBM Research Europe – Zurich, Rüschlikon, Switzerland

[2] Photonics Laboratory, ETH Zürich, Zürich, Switzerland, Switzerland

[3] AMO GmbH, Aachen, Germany

[4] Chair of Electronic Devices, RWTH Aachen University, Aachen, Germany

[5] Macromolecular Chemistry Group and Wuppertal Center for Smart Materials & Systems (CM@S), Bergische Universität Wuppertal, Wuppertal, Germany

[+] These authors contributed equally to this work.

[#] These authors have jointly supervised this work.

[*] Corresponding authors: dar@zurich.ibm.com, tof@zurich.ibm.com

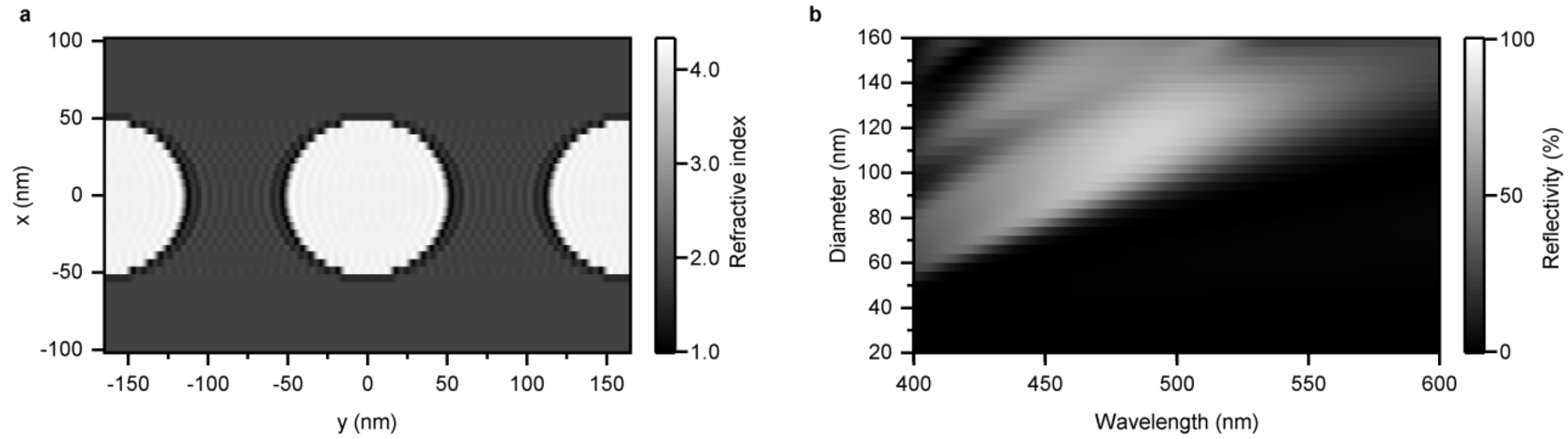

**Supplementary Fig. 1 | High contrast grating mirrors. a**, Map showing the spatial distribution of the refractive index along a silicon–MeLPPP HCG as used with the RCWA simulation. **b**, Reflectivity map of a HCG mirror as function of wavelength and diameter of the silicon pillars constituting the grating.

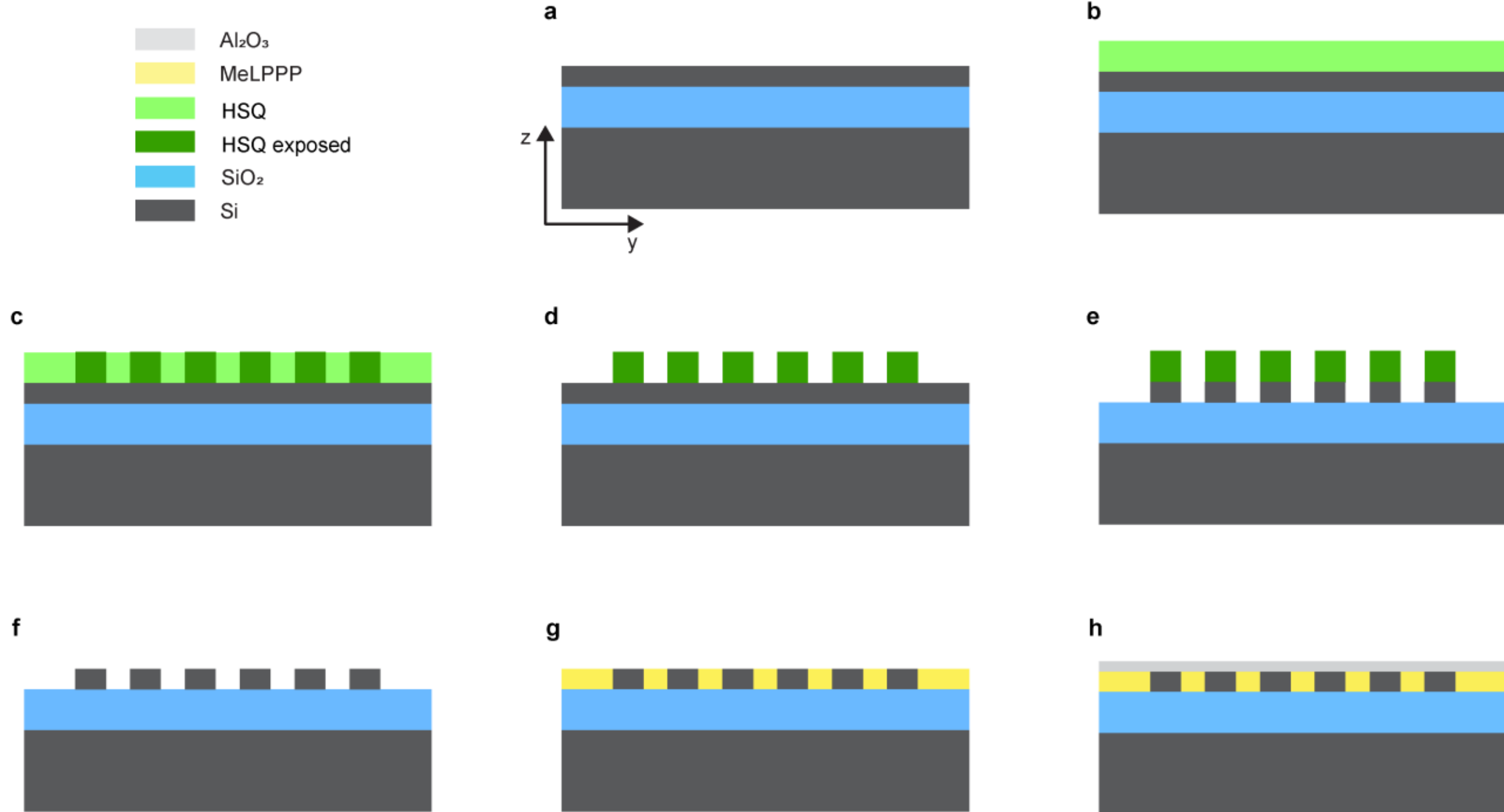

**Supplementary Fig. 2 | Fabrication process of high contrast grating microcavities. a**, SOI wafer with top silicon layer of 220 nm and 3 µm thick silica layer. **b**, Spin-coating deposition of HSQ resist. **c**, Exposure and patterning of HSQ with electron beam lithography. **d**, Development of HSQ. **e**, Etching of uncovered silicon. **f**, Removal of residual HSQ. **g**, Spin-coating deposition of MeLPPP. **h**, Encapsulation with alumina through electron beam deposition.

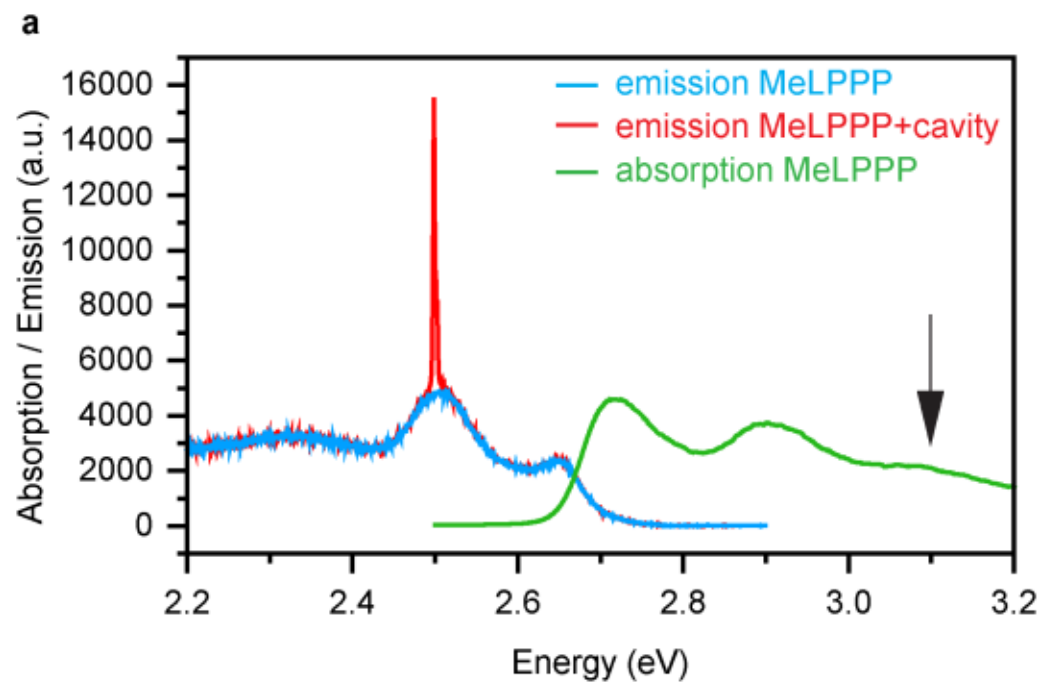

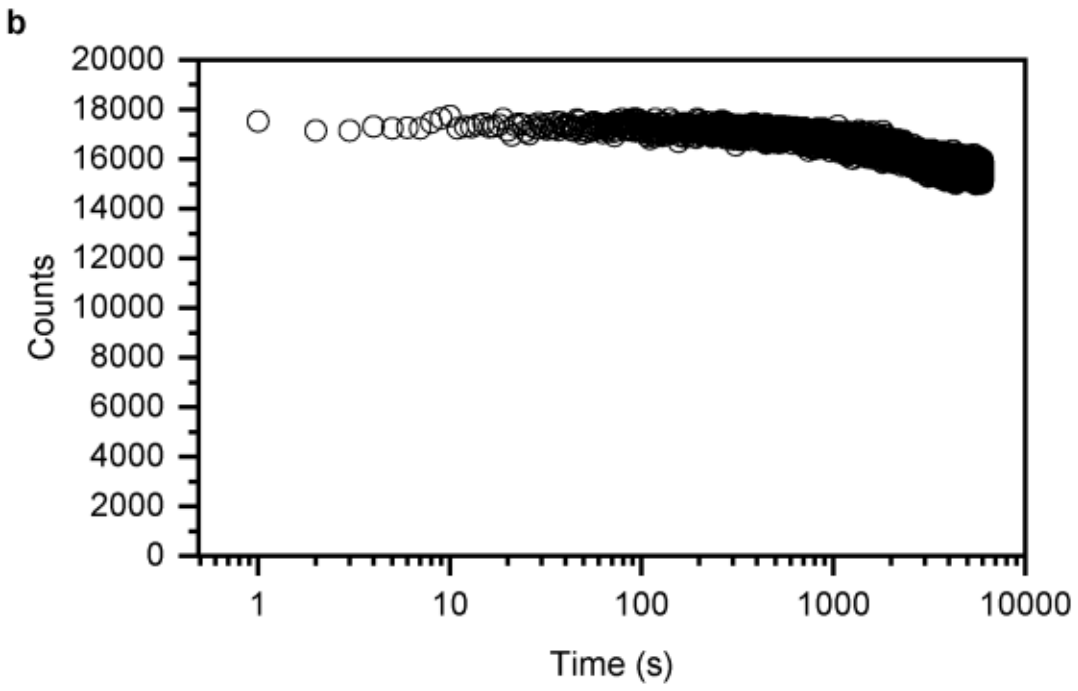

**Supplementary Fig. 3 | MeLPPP spectrum and photostability. a**, Photoluminescence spectra of MeLPPP measured by exciting at 3.1 eV (black arrow) with ~1.5 $P_{\mathrm{th}}$ on top of a HCG cavity (red) and outside of a HCG cavity (blue) and absorption spectrum (green). **b**, Spectrally integrated photoluminescence intensity measured outside of the HCG cavity as a function of time, excited at a fluence of ~1.5 $P_{\mathrm{th}}$ using 150 fs pulses with 1 kHz repetition rate.

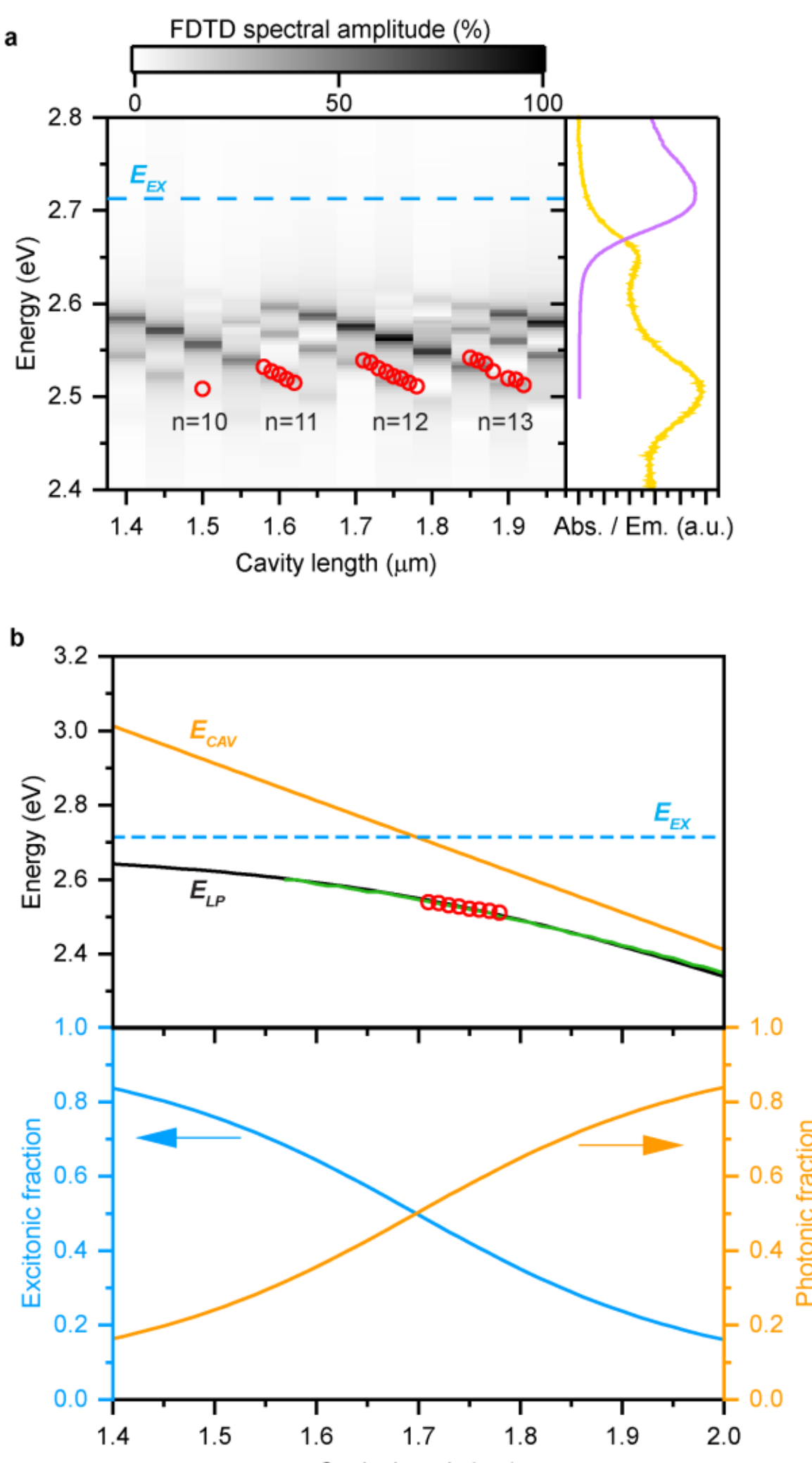

**Supplementary Fig. 4 | Simulations and coupled-oscillators model. a**, Cavity detuning plot (energy versus cavity length) of experimentally measured polariton condensation energies (red circles) and simulated spectral resonances (grey scale), designated by their longitudinal order *n*, as obtained from 3D FDTD calculations of HCG cavities in the strong light-matter interaction regime. The ab-initio simulations use the as-designed geometrical parameters and the measured refractive index dispersion from MeLPPP without any adjustments made to match to the experimental resonances. The exciton energy is represented by the dashed blue line. The right panel shows the absorption (purple) and emission (yellow) spectra of MeLPPP. **b**, Top panel shows the experimentally measured resonance energies of the longitudinal mode with $n$ = 12 (red circles) and a fit with a coupled-oscillators model showing the lower polariton branch (black). Cavity resonance energies displaying the strong light-matter interaction regime (green) and the weak-coupling regime (orange) have been extracted from RCWA simulations versus cavity length. The bottom panel shows the exciton (blue) and photon (orange) fraction, i.e. Hopfield coefficients, of the lower polariton branch versus the cavity length for this mode.

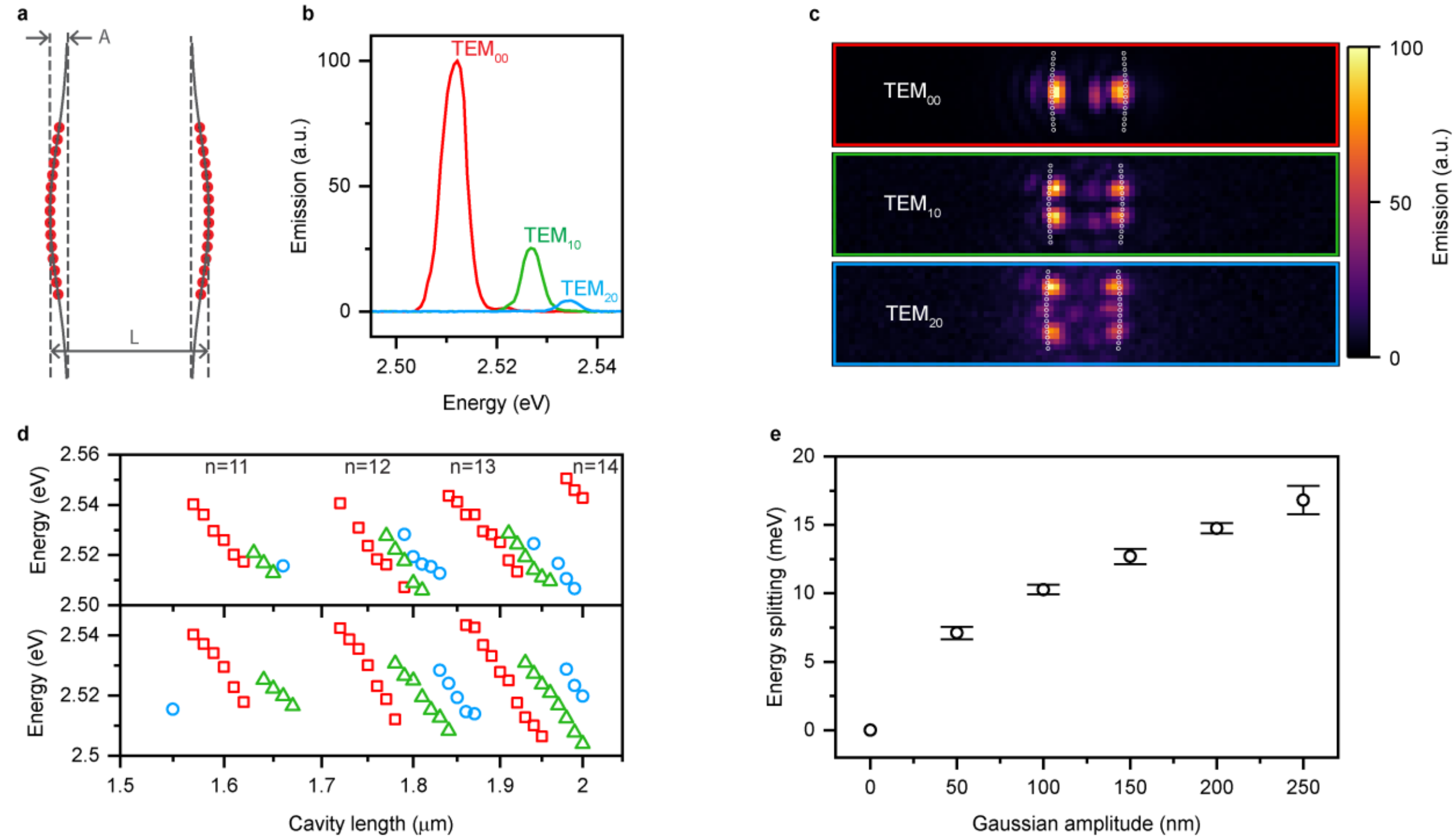

**Supplementary Fig. 5 | Engineering the high contrast grating cavity modes.** **a**, Scheme of a HCG cavity, highlighting the Gaussian curvature of the mirrors. **b**, Three energy-filtered emission spectra from a HCG cavity with cavity length $L$ = 1.92 μm and curvature described by a Gaussian amplitude of $A$ = 100 nm and a FWHM of 2.36 μm. The spectra relate to modes of different transversal order: $TEM_{00}$ (red), $TEM_{10}$ (green), $TEM_{20}$ (blue). **c**, Real-space images displaying the energy-filtered emission for the three modes shown in **a** with their characteristic scattering pattern on the HCG mirrors that reflects the modal profile in the cavity. **d**, Energy of cavity resonances as a function of cavity length for Gaussian-shaped cavities with $A$ = 100 nm (top panel) and $A$ = 250 nm (bottom panel). Different transversal orders are highlighted by different colors and symbols with $TEM_{00}$ (red), $TEM_{10}$ (green), $TEM_{20}$ (blue). **e**, Energy splitting between $TEM_{00}$ and $TEM_{10}$ as a function of Gaussian amplitude $A$. For the flat cavities ($A$ = 0 nm), there is no splitting as there are no discrete transversal orders. Vertical bars indicate the standard deviations of the energy splitting resulting from measurements on different devices having different cavity lengths.

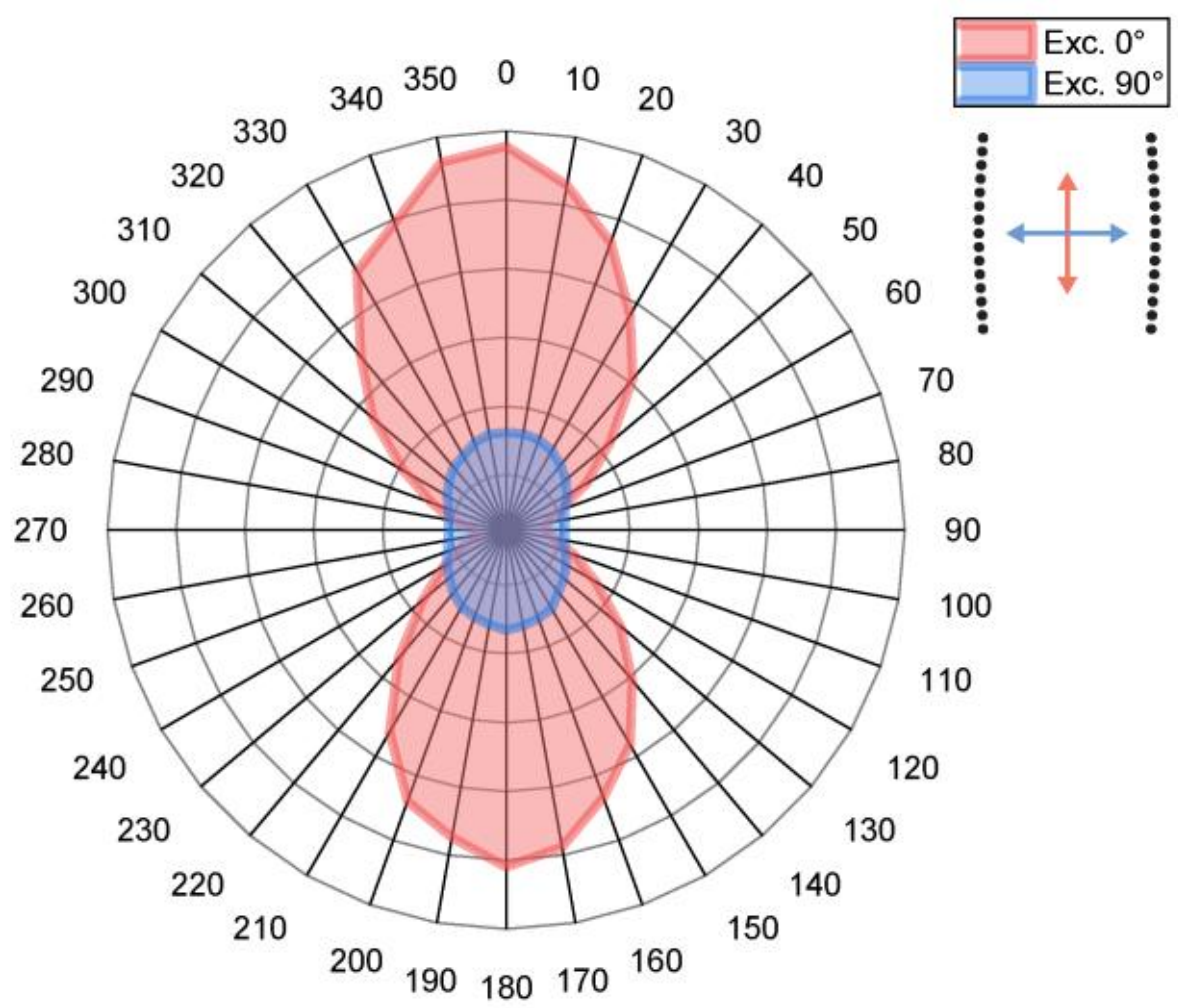

**Supplementary Fig. 6 | Polarization of the emission.** Emission intensity of the HCG cavity is shown as a function of angle of a polarizer that is placed in the detection path. When the excitation polarization is aligned parallel to the HCG gratings (red), polariton condensation occurs in the cavity with a mode polarization that is parallel to the HCG gratings. When excited with the same fluence but with orthogonal polarization (blue), no polarition condensation takes place, and the emission shows an almost angularly invariant pattern, typical for the unpolarized photoluminescence from a MeLPPP film, presumably slightly modified by the scattering from the HCG gratings.

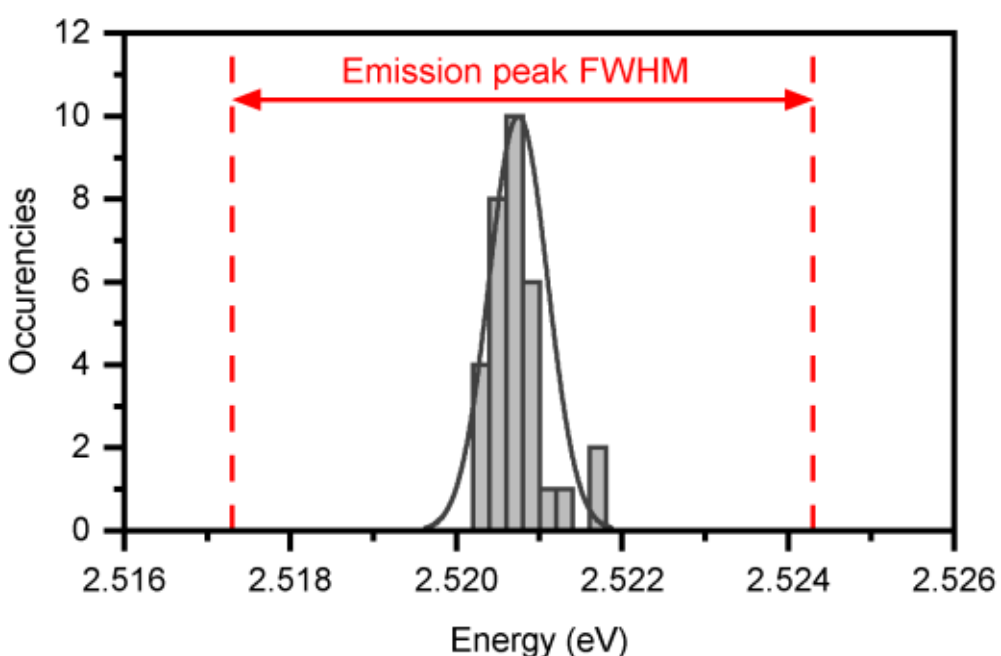

**Supplementary Fig. 7 | Statistics on high contrast grating cavities.** The distribution of the measured energy of the center of the condensation peaks of 32 HCG microcavities, which are identical by design (*L* = 1.91 µm), is represented with the gray histogram with bin size of 0.2 meV and is fitted with a normal distribution (black solid line) having a standard deviation of 0.47 meV. For comparison, the FWHM of the cavity mode obtained from the emission below threshold is shown in red (~ 7 meV).

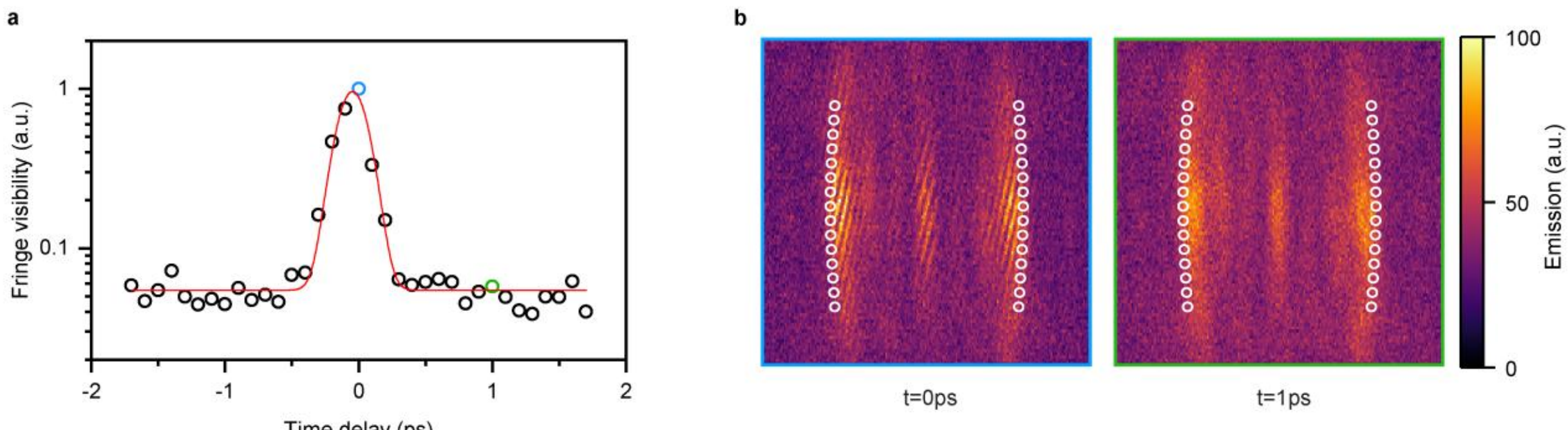

**Supplementary Fig. 8 | Temporal and spatial first-order coherence of the polariton condensate. a**, Michelson interference fringe visibility of the polariton condensate emission as a function of time delay between the interferometer arms, as obtained from Fourier transformation of the interferograms. The red line is a Gaussian fit to the data. **b**, Interferometric real-space images at 0 ps and 1 ps time delay, corresponding to the blue and green data points in **a**, respectively.

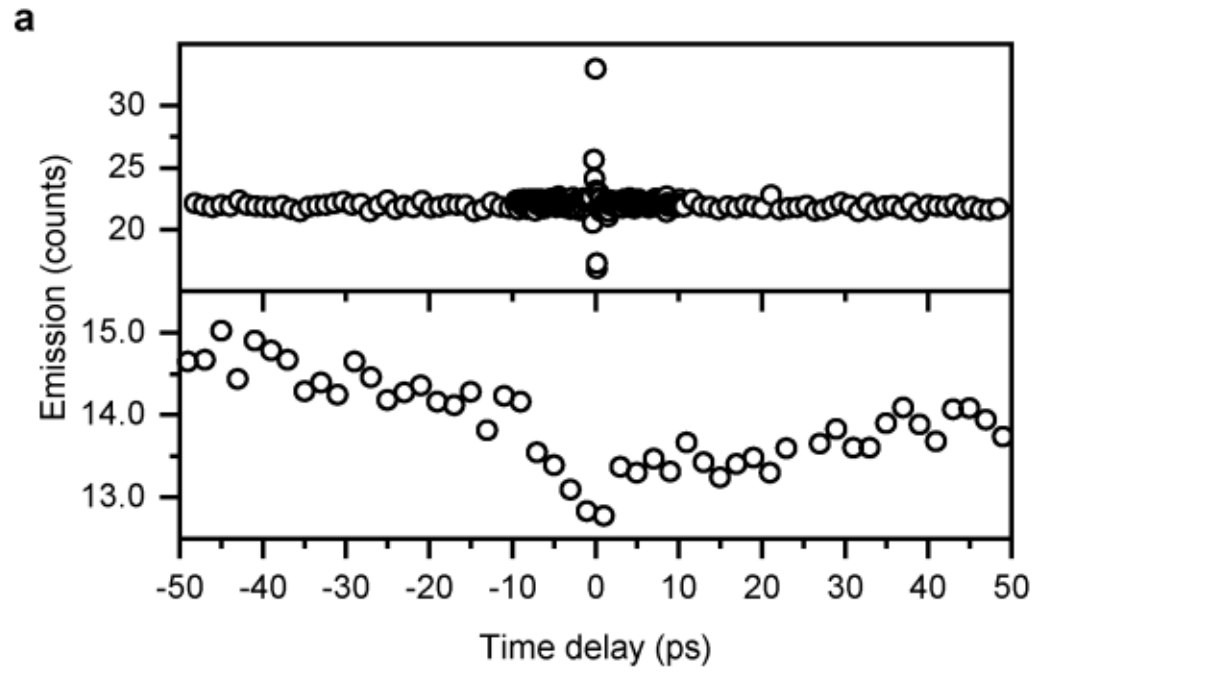

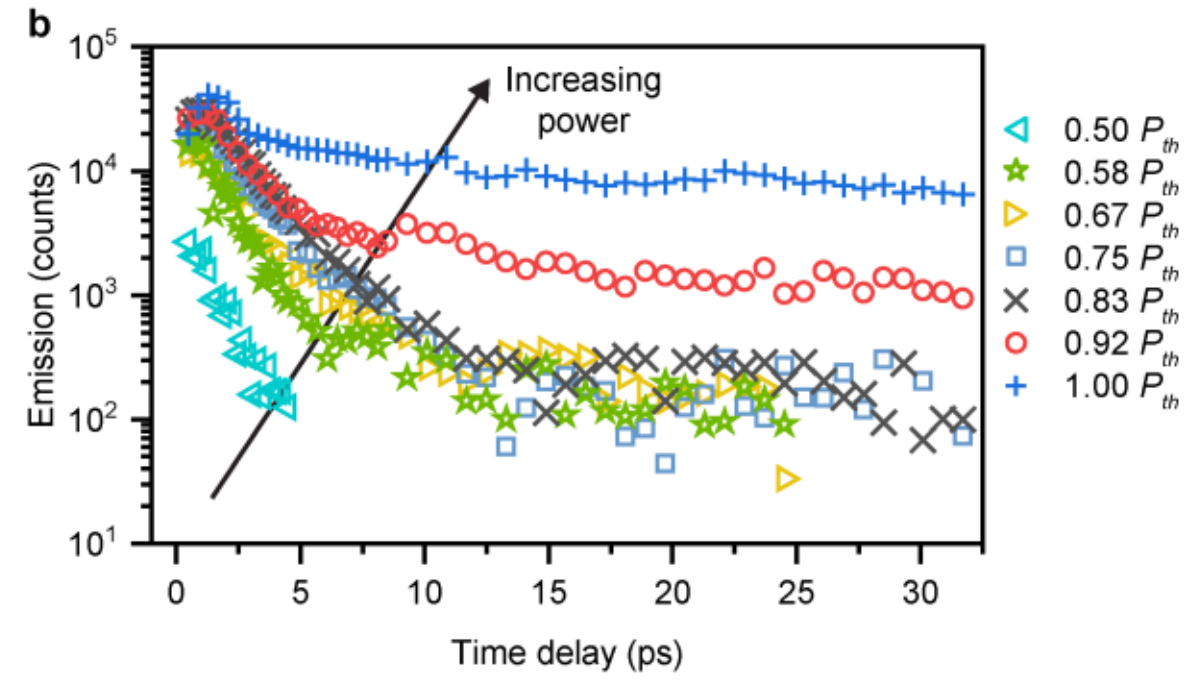

**Supplementary Fig. 9 | Time delay calibration and polariton condensation dynamics. a,** The top panel shows the intensity versus time delay for the reflections of two overlapping excitation beams (~150 fs duration), where interference is observed around 0 ps, that is then used as zero calibration point for the dynamical measurements. The bottom panel shows the photoluminescence intensity versus time delay between two overlapping excitation beams for an MeLPPP film without cavity, where the intensity shows a minimum around 0 ps due to nonlinear quenching from exciton annihilation. Notably, the quenching reduces with a ~22 ps time constant, in line with the exciton lifetime for this material. **b**, Integrated intensity of emission spectra of the condensate resulting from the excitation of two spatially overlapping beams on a single HCG cavity as function of time delay for different excitation power (same fluence is used for both beams).

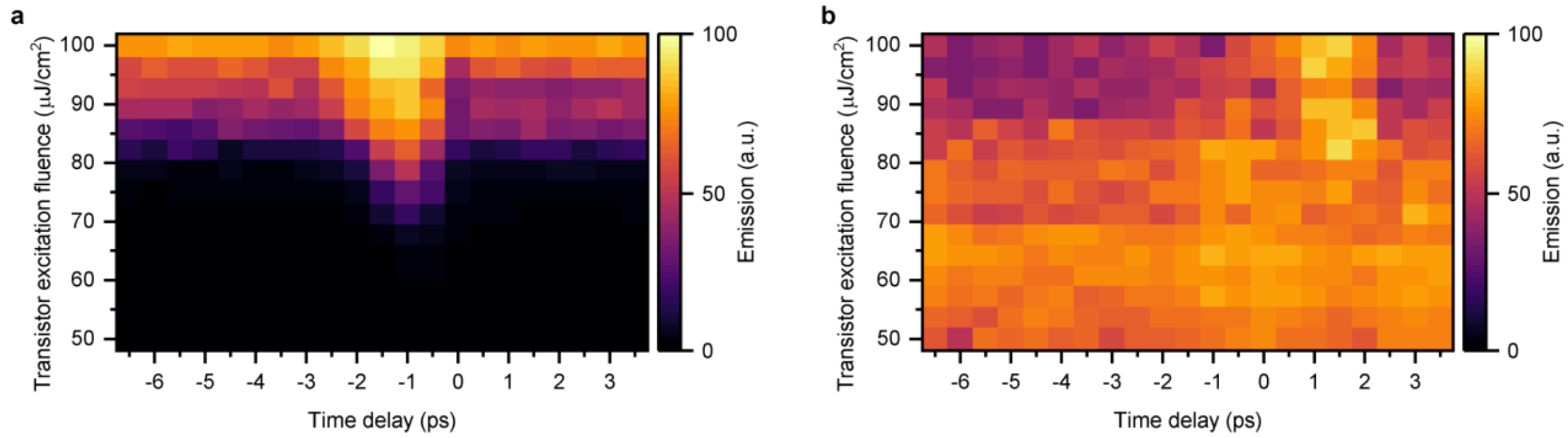

**Supplementary Fig. 10 | Reversing the roles of seed and transistor cavity. a**, Real-space emission intensity integrated over the transistor cavity area in the emission image versus transistor excitation fluence and time delay between seed ($P_{\mathrm{seed}} = 1.5\ P_{\mathrm{th}}$) and transistor excitation, showing a maximum around -1 ps time delay. This analysis of the image intensities yields to almost identical results as the spectral analysis of Fig. 3c in the main manuscript but provides the possibility to analyze simultaneously different structures. **b**, Same experiment as **a**, but here the emission intensity is integrated over the seed cavity area, showing a maximum near +1 ps time delay.